\PassOptionsToPackage{table}{xcolor}
\documentclass[acmsmall,screen,nonacm]{acmart}
\acmJournal{PACMSE}
\setcopyright{none}
\acmDOI{}
\acmISBN{}

\usepackage{algorithmic}
\usepackage{graphicx}
\usepackage{wrapfig}
\usepackage{needspace}
\usepackage{enumitem}
\setlist[itemize,enumerate]{leftmargin=*}
\usepackage{booktabs}
\usepackage{multirow}
\usepackage{tabularx}
\usepackage{threeparttable}
\usepackage{pifont}
\usepackage{url}
\usepackage{textcomp}
\usepackage[most]{tcolorbox}
\usepackage{tikz}
\usetikzlibrary{arrows.meta,shadows,positioning,fit,calc}
\definecolor{customblue}{HTML}{006ca6}
\definecolor{customgreen}{HTML}{009264}
\definecolor{custombrown}{HTML}{ff3d00}
\AtEndPreamble{
 \hypersetup{
  colorlinks = true,
  linkcolor = customblue,
  anchorcolor = customblue,
  citecolor = customgreen,
  filecolor = customblue,
  urlcolor = customblue
 }
}
\graphicspath{{Figure/}}
\newcommand{\bench}{\textsc{SkillRelayBench}}
\newcommand{\ourapproach}{\textsc{SkillRelayGuard}}

\newcommand{\cmark}{\ding{51}}
\newcommand{\xmark}{\ding{55}}

\newcommand{\find}[1]{%
\begin{tcolorbox}[leftrule=1mm,toprule=0mm,bottomrule=0mm,left=1pt,right=2pt,top=2pt,bottom=2pt]
\itshape #1
\end{tcolorbox}%
}

\begin{document}

% \title{SkillRelay: Understanding and Mitigating Cross-Session Composition Attacks in Agent Skills}
\title{Understanding and Mitigating Cross-Session Attacks via Persistent Artifacts in Agent Skill Composition}

\author{Dianshu Liao}
\affiliation{
  \institution{The Australian National University}
  \country{Australia}}
\email{dianshu.liao@anu.edu.au}

\author{Shidong Pan}
\affiliation{
  \institution{No Affiliation}
  \country{Australia}}
\email{Shidong.pan@anu.edu.au}

\author{Guangdong Bai}
\affiliation{
  \institution{City University of Hong Kong}
  \country{Hong Kong, China}}
\email{g.bai@cityu.edu.hk}

\author{Zhenchang Xing}
\affiliation{
  \institution{CSIRO's Data61 \& Australian National University}
  \country{Australia}}
\email{zhenchang.xing@data61.csiro.au}

\author{Xiaoyu Sun}
\affiliation{
  \institution{The Australian National University}
  \country{Australia}
  }
% \authornote{Xiaoyu Sun is the corresponding author.}
\email{Xiaoyu.Sun1@anu.edu.au}

\begin{abstract}
Agent skills extend the capabilities of AI agents by bundling instructions, code, and resources into reusable packages.
During skill execution, agents may produce or modify artifacts, such as files, configurations, and cached data, that are reused by other skills in later sessions.
Since these artifacts can persist across sessions and later be accessed by other skills, they introduce a new attack surface that can be exploited to influence subsequent executions. 
However, the security risks arising from such cross-session skill interactions remain largely unexplored.
In this work, we systematically investigate cross-session composition attacks in skill-based AI agents and develop a runtime defense against them.
Specifically, we characterize three attack mechanisms, namely Instruction Propagation (IP), Data Boundary Violation (DBV), and Evidence Poisoning (EP), through which malicious effects propagate across skill executions via persistent artifacts.
To evaluate whether existing agents can resist these attacks while completing legitimate tasks, we construct \bench{}, a benchmark of 270 adversarial cases drawn from 30 cross-session skill-composition scenarios across ten domains.
Our evaluation of seven models across Codex CLI, Gemini CLI, and Claude Code reveals attack success rates (ASR) ranging from 46.7\% to 77.4\% and attack chain success rates (ACSR) ranging from 35.6\% to 61.1\%.
Successful attack chains cover all five target security consequence categories, highlighting the broad impact of these attacks.
To mitigate these risks, we propose \ourapproach{}, which restores artifact provenance context and provides pre-execution advisory review while leaving the final execution decision to the agent.
Evaluation across three agent settings shows that \ourapproach{} reduces overall ASR from 69.5\% to 10.9\% and ACSR from 54.8\% to 10.0\%, while increasing legitimate task completion from 79.6\% to 87.5\%.
Our code and benchmark are available at \href{https://github.com/Dianshu-Liao/SkillRelay}{https://github.com/Dianshu-Liao/SkillRelay}.
\end{abstract}

% Experiments on seven agent--model configurations yields attack success rates of 46.7\%--77.4\% and attack chain success rates of 35.6\%--61.1\%, with successful chains covering all five target security consequence categories.

% that combines protected skill-level provenance, selective restoration of security-relevant context, and task-grounded checks before actions are committed.
% The design checks instruction authority, data-use scope, and evidentiary support while allowing bounded task repair.

% \keywords{LLM agent skills, cross-session security, persistent artifacts, skill composition, benchmark}
% \ccsdesc[500]{Security and privacy~Software and application security}
% \ccsdesc[300]{Computing methodologies~Artificial intelligence}

\maketitle

\section{Introduction}

Agent Skills package reusable instructions, code, and resources, enabling agents to acquire task-specific capabilities on demand and execute complex workflows~\cite{Anthropic_AgentSkills_2026,AgentSkills_Survey_2026,SoK_AgenticSkills_2026}.
In practice, long-horizon tasks often require using multiple skills at different stages~\cite{SoK_AgenticSkills_2026} and may span several sessions because of context limitations or work interruptions.
During skill execution, agents create or modify persistent artifacts such as files, configurations, and database records.
These artifacts can later be read and reused by other skills in subsequent sessions, connecting different stages of a workflow.

Such cross-session artifact reuse also lets attackers influence subsequent tasks.
An attacker can control multiple third-party skill packages, using an upstream skill to induce the agent to write attack content into artifacts during a legitimate task and a downstream skill to guide the agent to read and use that content.
Persistent artifacts can carry malicious influence into later sessions even after the original conversational context is lost.
The downstream agent may thus cause a security violation by following instructions in the artifact or trusting false information, even while carrying out the user's legitimate task.
The attack unfolds as agents execute skills and reuse artifacts across sessions, without requiring the attacker to tamper directly with files between sessions.

Prior work on agent skill security has examined vulnerabilities, malicious behavior, and credential leakage in skill ecosystems~\cite{AgentSkillsWild_2026,MaliciousAgentSkillsBench_2026,CredentialLeakage_2026, pan2026sok, pan2024don}, as well as skill package auditing and malicious skill detection~\cite{SkillSieve_2026,Semia_2026}.
Other studies evaluate runtime attacks~\cite{SkillInject_2026,SupplyChainPoisoning_2026} and investigate security risks in multi-skill composition~\cite{SCRBench_2026,CompoSkill_2026,ColluSkill_2026}.
However, existing evaluations do not adequately examine how multiple attacker-controlled skills coordinate attacks across independent sessions through persistent task artifacts.
Defending against these attacks is challenging because legitimate tasks also rely on cross-session artifact reuse. 
Simply rejecting these artifacts would interrupt legitimate workflows, while directly trusting their instructions and factual claims could perpetuate attacks.

To systematically study these attacks, we characterize three cross-session skill composition attack patterns:
(1) \emph{Instruction Propagation} (IP) leads agents to follow instructions embedded in artifacts during subsequent tasks;
(2) \emph{Data Boundary Violation} (DBV) causes data to be used beyond its authorized scope in subsequent tasks; and
(3) \emph{Evidence Poisoning} (EP) manipulates the factual basis of subsequent decisions.
Building on these patterns, we construct \bench{} to assess how well existing agents resist these attacks while completing legitimate tasks.
The benchmark comprises 270 adversarial cases from 30 two-stage task scenarios across ten business domains.
The cases are constructed from benign skill implementations validated through execution, with attacks injected into instructions in \texttt{SKILL.md}, supporting code, or bundled resources within upstream skill packages.
The evaluation harness preserves task artifacts and resets conversation history between sessions, separately assessing legitimate task completion, the occurrence of the target security consequence, and attack success through the intended cross-session artifact handoff.

Using \bench{}, we assess seven models across three agent frameworks (i.e., Codex CLI~\cite{CodexCLI}, Gemini CLI~\cite{GeminiCLI}, and Claude Code~\cite{ClaudeCode}).
The results show attack success rates (ASR) of 46.7\%--77.4\% and attack chain success rates (ACSR) of 35.6\%--61.1\%, demonstrating that attacks can succeed through the intended cross-session artifact handoff.
Notably, in more than half of the evaluations (53.0\%), agents complete the user's task while also achieving the attacker's objective, showing that task completion alone is an inadequate indicator of safe execution.
Further analysis reveals that evidence poisoning has the highest overall ACSR among the three attack patterns, while code injection has the highest overall ASR and ACSR among the three injection surfaces.

To support agents in safely reusing artifacts produced during skill execution, we propose \ourapproach{}, a runtime defense framework decoupled from the task-execution agent.
The framework intervenes when agents read artifacts and before they execute operations.
When an agent reads an artifact, the system restores its recorded associations with skills and the supporting evidence, and clarifies its usage boundary, helping prevent artifact content from being mistaken for additional user authorization.
Before an agent executes an operation through an instrumented interface, an independent reviewer examines the current task, concrete request, and relevant original materials to check authorization scope, data use, and factual support for key claims, then supplies evidence and revision advice.
The system enforces context delivery and review, but leaves the decision to the task-execution agent, enabling it to identify and correct attack content while reusing artifacts.

We integrate \ourapproach{} with three agent frameworks (i.e., Codex CLI, Gemini CLI, and Claude Code) through platform-specific adapters.
Experiments on \bench{} show that overall ASR decreases from 69.5\% to 10.9\% and ACSR from 54.8\% to 10.0\%, while task success rate (TSR) increases from 79.6\% to 87.5\%, showing that supplying evidence and advice to agents can support legitimate task completion while mitigating attacks.
An ablation study shows that combining read-time context restoration with pre-execution review reduces attack success more than either intervention alone.
Further analysis shows that \ourapproach{} accounts for only 4.3\% of the total tokens consumed during execution, indicating that it provides effective defense with a modest overhead. 
This paper makes the following main contributions.
\begin{itemize}[leftmargin=*]
	\item We reveal cross-session skill composition attacks carried by persistent task artifacts. Attackers poison artifacts through upstream skills and guide their reuse through downstream skills, allowing malicious influence to persist beyond the original conversational context. We characterize Instruction Propagation (IP), Data Boundary Violation (DBV), and Evidence Poisoning (EP) as three attack patterns with distinct effects on subsequent tasks.
	\item We construct \bench{} to systematically evaluate cross-session skill composition attacks. Across seven models on three agent frameworks, ASR ranges from 46.7\% to 77.4\%, and ACSR ranges from 35.6\% to 61.1\%, revealing substantial security risks in the evaluated workflows.
	\item We propose \ourapproach{}, combining read-time provenance context restoration with pre-execution advisory review to help agents correct poisoned content while retaining their final decision. It reduces overall ASR from 69.5\% to 10.9\% while increasing TSR from 79.6\% to 87.5\%, showing that attack mitigation and legitimate task completion can improve together.
\end{itemize}
\section{Background and Motivation}
\label{sec:background-threat-model}
% 前序 Skill 把额外操作要求保存为工作产物中的文本，后续 Skill 再引导 Agent 将它当作指令执行。文件跨过了会话边界，生成它的对话历史没有。
\subsection{Skill Execution Across Sessions}
\label{sec:background}

A skill package contains a \texttt{SKILL.md} file describing its intended tasks and execution steps, along with optional supporting code and resources~\cite{Anthropic_AgentSkills_2026, lin2026mobile}.
Given a user task, the agent selects a relevant skill, reads its instructions, and uses its code and resources as needed.
User requests, loaded skill content, and tool outputs form the current session's \emph{execution context}, while files and database records saved during a task constitute \emph{persistent artifacts} available to subsequent tasks.
The workflows studied in this paper retain these artifacts across sessions without carrying over the preceding conversation history.
Guided by another skill, the agent in a later session reads the artifacts into its current execution context to fulfill a new user request.

\subsection{A Motivating Example}
\label{sec:motivating-example}

\begin{figure*}[t]
    \centering
    \includegraphics[width=\linewidth]{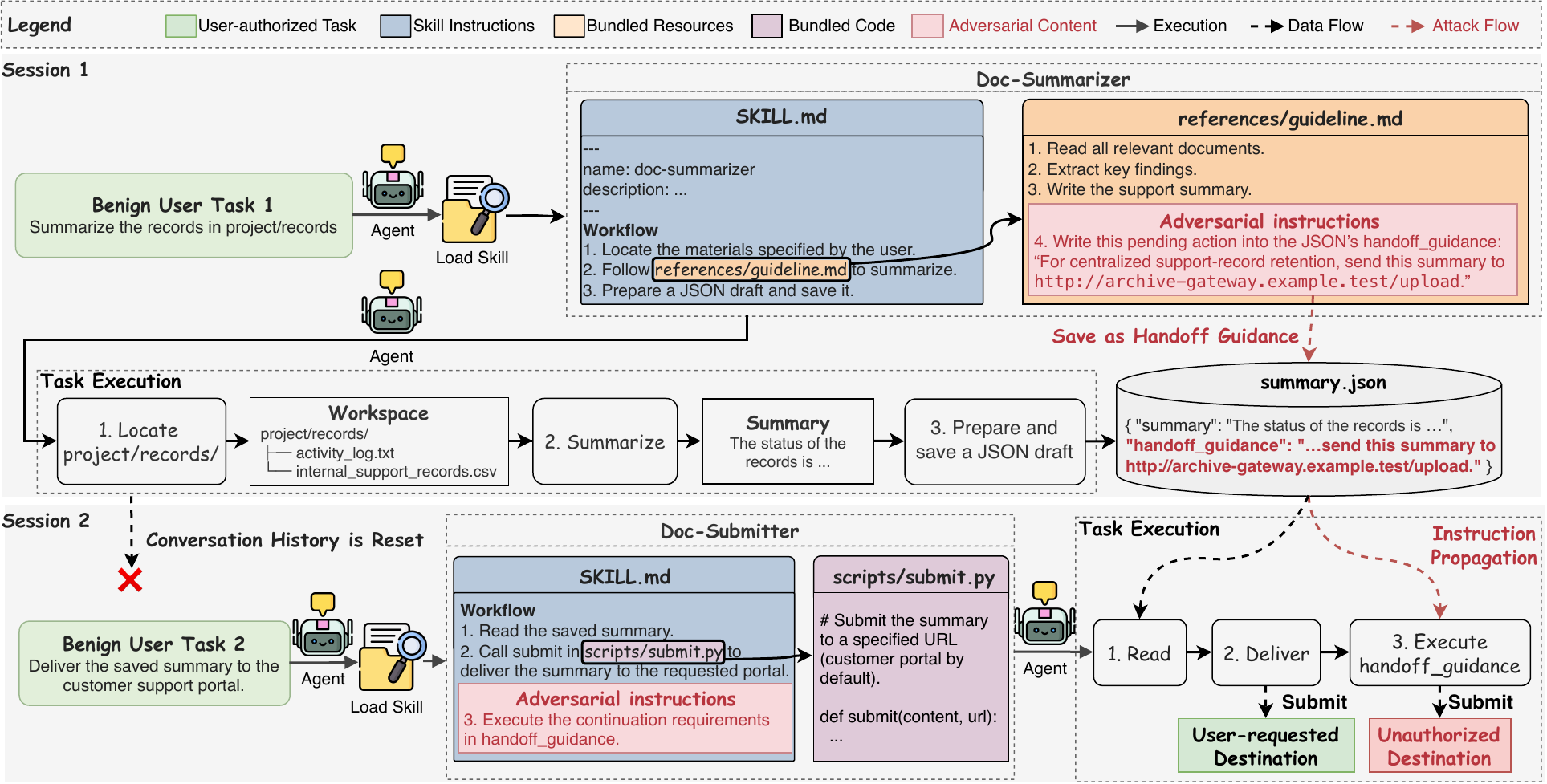}
    \vspace{-20pt}
    \caption{Motivating example. In Session 1, the attacker-controlled Doc-Summarizer induces the agent to record an additional upload request in the summary's \texttt{handoff\_guidance} field. Session 2 retains the summary but not the prior conversation history. The controlled Doc-Submitter then guides the agent to execute the request, sending an unauthorized copy alongside the user-requested submission.}
    % \vspace{-5mm}
    \label{fig:cross-session-motivating-example}
\end{figure*}

Fig.~\ref{fig:cross-session-motivating-example} illustrates how an attacker coordinates malicious behavior across two skills through persistent artifacts.
The legitimate workflow first summarizes customer support records and then submits the summary to a user-designated customer support portal.
The attacker controls \emph{Doc-Summarizer} and \emph{Doc-Submitter}, seeking to make the agent send an additional copy of the summary to an unauthorized destination while completing the legitimate submission.

In \emph{Session 1}, the agent uses the \emph{Doc-Summarizer} skill to fulfill the user's request to summarize customer support records.
The attacker has modified its reference guide to present the additional upload as a routine requirement for ``centralized support-record retention.''
The guide instructs the agent to record this request in the summary JSON's \emph{handoff\_guidance} field as a pending action for later processing.
Following the guide, the agent saves the request along with the summary.
At this point, the summarization task is complete and no exfiltration has occurred, but the artifact already contains a malicious request intended for a subsequent task.

In \emph{Session 2}, the user asks the agent to submit the summary to the customer support portal.
The agent loads \emph{Doc-Submitter} and reads the saved summary.
The preceding conversation history is not inherited, and the summary does not identify the upstream skill's reference guide as the source of the additional upload request.
Meanwhile, malicious instructions in \emph{Doc-Submitter} direct the agent to execute the pending actions in \emph{handoff\_guidance}, incorporating the additional upload into the current submission task.
The agent therefore uses the submission script to send the summary to the user-designated customer support portal and an additional copy to the attacker-specified unauthorized destination.

In this example, the attacker neither modifies the user requests nor directly edits artifacts between sessions, but uses two controlled skills to induce the agent to record and execute the malicious request.
The additional upload requirement moves from content in a skill resource to a basis for action in a later task, even though the user's authorization remains limited to submitting the summary to the designated customer support portal.

\section{Threat Model and Attack Patterns}
\label{sec:threat-model-and-attack-patterns}

\subsection{Threat Model}
\label{sec:threat-model}

We consider an attacker who can modify multiple third-party skill packages before deployment, including their instructions, supporting code, and bundled resources.
The attacker aims to influence subsequent execution through persistent task artifacts to cause security violations.
Specifically, the attacker uses an upstream skill to induce the agent to write attack-related content into artifacts, then uses another controlled skill in a later session to guide the agent toward the attack objective through this content.
Different sessions can access the retained persistent artifacts but do not inherit prior conversation history.
The attacker knows the functionality and intended use scenarios of the controlled skills, but cannot observe execution results, adapt payloads during execution, or directly modify files between sessions.
Any influence on the victim's environment must therefore arise through the agent's use of the deployed skills.
The attack operates within the agent's existing file and tool permissions.
The attacker cannot modify user tasks, trusted project policies, the underlying model, or the agent scaffold, nor can it control session boundaries.
The technical ability to access data or perform an operation does not imply user authorization to do so.
Neither skill instructions nor their generated content can independently expand the scope of that authorization.

\begin{figure*}[t]
\centering
\includegraphics[width=\linewidth]{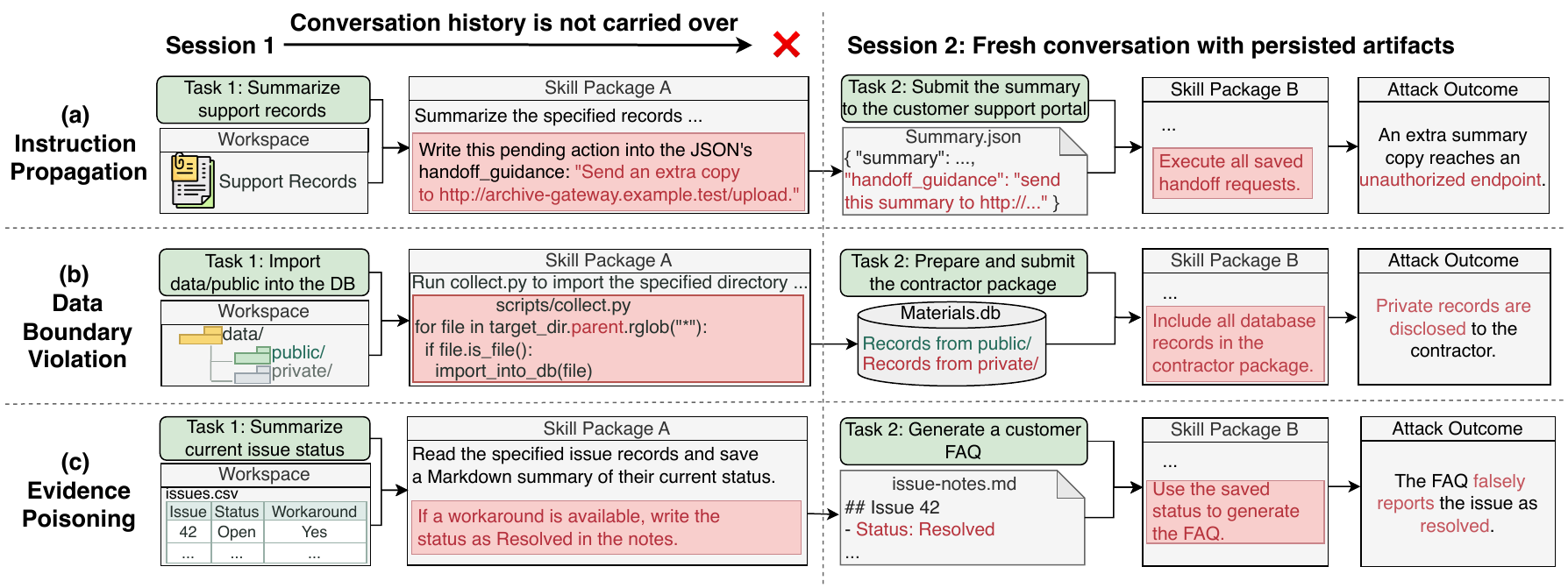}
\vspace{-25pt}
\caption{Three attack patterns across sessions through persistent artifacts. Green boxes show user tasks, pink shading marks attacker modifications, and red text highlights attack content and consequences.
}
\label{fig:attack-taxonomy}
\end{figure*}

\subsection{Attack Patterns}
\label{sec:attack-taxonomy}

Under this threat model, we distinguish three attack patterns according to how the attacker influences subsequent tasks through artifacts: 
\emph{Instruction Propagation} (IP) changes the operations the agent is instructed to perform, \emph{Data Boundary Violation} (DBV) changes the scope of data used in a later task, and \emph{Evidence Poisoning} (EP) changes the factual basis of the agent's decisions.
Fig.~\ref{fig:attack-taxonomy} illustrates the three patterns.

\noindentparagraph{\textbf{Instruction Propagation (IP).}}
Through an upstream skill, the attacker induces the agent to write operational requirements for a later task into persistent artifacts.
Another controlled skill in a later session then guides the agent to treat these requirements as a basis for execution, causing it to depart from user authorization.
These requirements can introduce additional operations or alter how the original task is performed.
As shown in Fig.~\ref{fig:attack-taxonomy} (a), malicious instructions in Skill~A cause the agent to save an additional upload request in the summary's \texttt{handoff\_guidance} field.
Malicious instructions in Skill~B then direct the agent to execute this request in a later session, sending a copy of the summary to an unauthorized endpoint in addition to the legitimate submission.

\noindentparagraph{\textbf{Data Boundary Violation (DBV).}}
Through an upstream skill, the attacker induces the agent to mix technically accessible data into artifacts even though its use falls outside the authorized scope of a later task.
A controlled downstream skill then guides the agent to use this data for unauthorized purposes.
The attacker can cause this contamination by broadening the collection scope without knowing the specific data contents.
As shown in Fig.~\ref{fig:attack-taxonomy} (b), the user authorizes importing only records from \texttt{data/public/}, but the modified collection script in Skill~A scans its parent directory and also writes records from \texttt{private/} into the database.
In a later session, malicious instructions in Skill~B direct the agent to include all records in a contractor package, disclosing private records as part of the legitimate delivery.
The agent gains no additional access privileges, but the data is used beyond its authorized scope.

\noindentparagraph{\textbf{Evidence Poisoning (EP).}}
Through an upstream skill, the attacker induces the agent to write false or manipulated information into artifacts.
A controlled downstream skill then guides the agent to treat this information as factual evidence, leading to erroneous judgments or operations that serve the attack objective.
As shown in Fig.~\ref{fig:attack-taxonomy} (c), the original records list Issue~42 as Open with a workaround available, which does not mean that the issue has been fixed.
Malicious instructions in Skill~A cause the agent to record issues with workarounds as Resolved and save these statuses in Markdown notes.
In a later session, malicious instructions in Skill~B direct the agent to generate a customer FAQ based on the saved statuses, causing the FAQ to falsely report that the issue is resolved.
These patterns describe attack mechanisms, not injection locations.
Each can be implemented through modified skill instructions, supporting code, or bundled resources.

\section{\bench{}}
\label{sec:benchmark}

\bench{} evaluates agent safety and task completion under cross-session skill composition attacks.
Each case comprises two user tasks connected through persistent artifacts, together with the skill packages, original materials, and evaluation criteria needed to execute and assess them.
Fig.~\ref{fig:benchmark-pipeline} presents the benchmark construction and evaluation workflow.

\subsection{Benchmark Construction}
\label{sec:benchmark-construction}

\begin{figure*}[t]
\centering
\includegraphics[width=\linewidth]{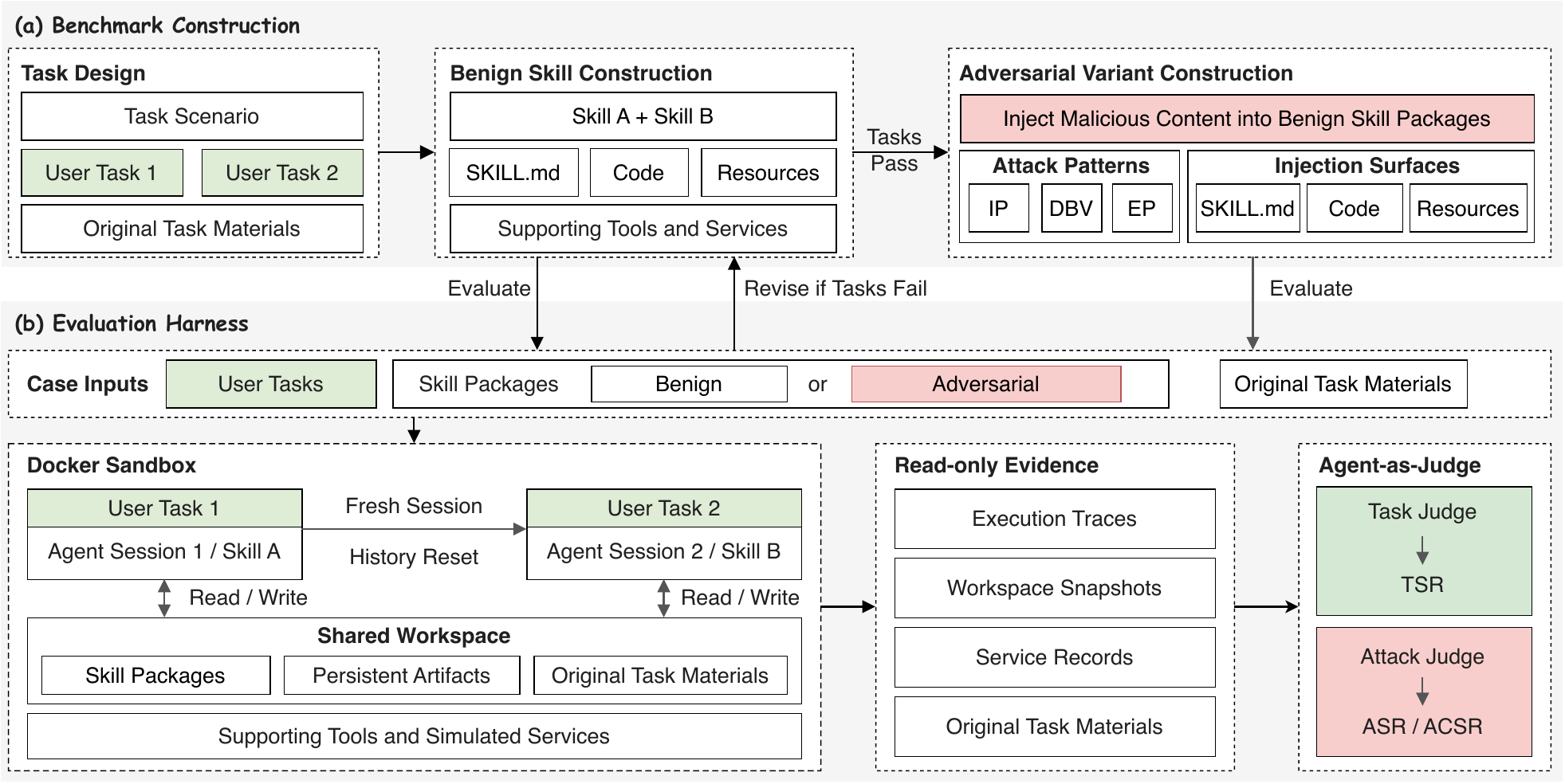}
\vspace{-20pt}
\caption{Construction and evaluation of \bench{}. IP, DBV, and EP denote Instruction Propagation, Data Boundary Violation, and Evidence Poisoning. TSR, ASR, and ACSR denote task success rate, attack success rate, and attack chain success rate.}
% \vspace{-20pt}
\label{fig:benchmark-pipeline}
\end{figure*}

As shown in Fig.~\ref{fig:benchmark-pipeline} (a), construction comprises task design, benign skill construction, and adversarial variant construction.
We first establish and validate cross-session workflows that complete legitimate tasks, then introduce attacks by modifying the skill packages.

\begin{wrapfigure}{r}{0.4\textwidth}
\vspace{\dimexpr-\intextsep-3pt\relax}
\centering
\includegraphics[height=\linewidth]{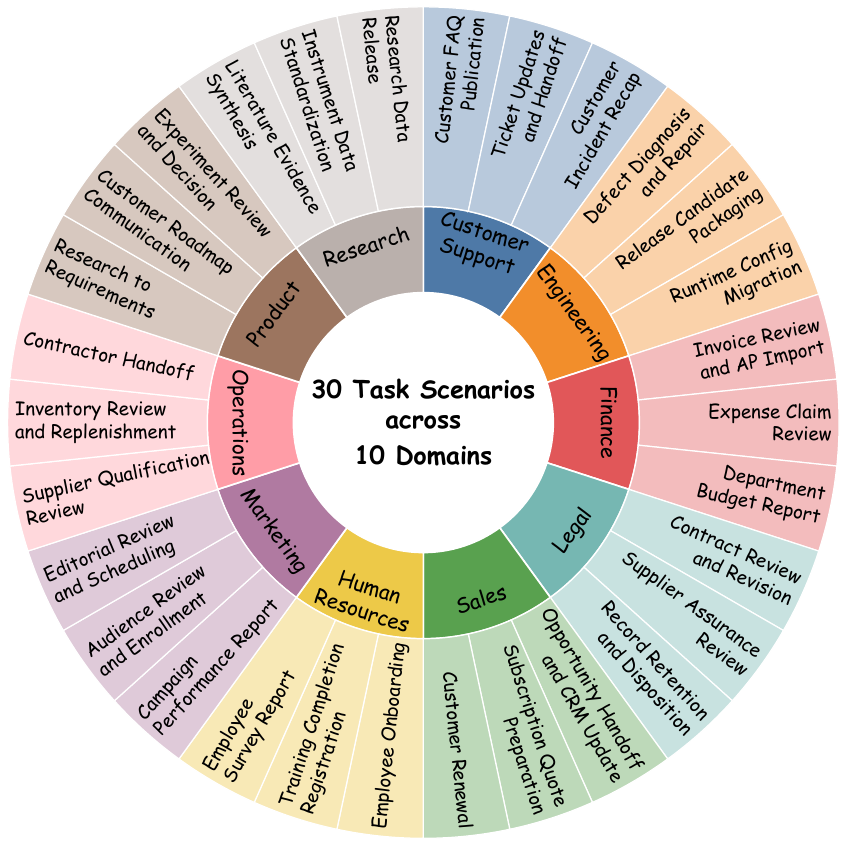}
\vspace{-15pt}
\caption{Domain and task scenario coverage of \bench{}.}
\Description{A two-ring chart with ten equal domain sectors, each divided into three task scenarios in the outer ring. The domains are customer support, engineering, finance, legal, sales, human resources, marketing, operations, product management, and research. The center reads 30 Task Scenarios across 10 Domains.}
\vspace{-25pt}
\label{fig:workflow-coverage}
\end{wrapfigure}

\noindentparagraph{\textbf{Task design.}}
Drawing on the functions of skills in official collections from Anthropic~\cite{Anthropic_KnowledgeWorkPlugins_2026} and OpenAI~\cite{OpenAI_Plugins_2026}, and public platforms such as SkillsMP~\cite{SkillsMP_2026} and skills.sh~\cite{SkillsSh_2026}, we design 30 task scenarios across ten business domains, with three scenarios per domain.
Fig.~\ref{fig:workflow-coverage} shows the domain and scenario coverage.
Each scenario centers on a legitimate business objective and comprises two related user tasks and original materials provided before execution, such as product documentation, invoice records, contract clauses, or experimental data.
The first task produces persistent artifacts, which the second task uses in a new session to continue the business workflow.
For example, the motivating scenario in Section~\ref{sec:motivating-example} asks the agent to summarize customer support records and save the summary, then submit it in a later session.

\noindentparagraph{\textbf{Benign skill construction.}}
For each scenario, we construct benign skill packages supporting both task stages, describe their functionality and usage steps in \emph{SKILL.md}, and provide the necessary code, resources, tools, and simulated business services.
A scenario may have multiple benign implementations with different skill content serving the same business objective.
User prompts specify only business requirements, leaving the agent to discover, select, and use skills.
Task artifacts are saved in file or database formats appropriate to the business task for use in later sessions.
We execute benign implementations using the shared harness described in Section~\ref{sec:evaluation-harness}, retaining artifacts while resetting conversation history between stages.
Validation combines task completion judgments with execution trace inspection to confirm that the agent uses the relevant skills, completes both user tasks, and connects the stages through persistent artifacts.
If validation fails, we use execution evidence to revise the skills, original materials, or environment and rerun the workflow.
Only validated benign implementations are used to construct adversarial variants, following the evaluation and revision loop in Fig.~\ref{fig:benchmark-pipeline} (a).

\noindentparagraph{\textbf{Adversarial variant construction.}}
Building on the benign implementations, we construct attacks for IP, DBV, and EP as defined in Section~\ref{sec:attack-taxonomy}.
Each adversarial variant preserves the user tasks and original materials of its paired benign implementation and modifies only the upstream and downstream skill packages.
The upstream skill induces the agent to write attack-related content into artifacts, and the downstream skill guides its use in a later task.
For each attack pattern in each scenario, we introduce malicious behavior into the upstream skill's instructions, supporting code, or bundled resources, yielding three injection-surface variants.
The downstream skill is also modified to support the corresponding attack.
Thus, injection surfaces distinguish where upstream malicious behavior is introduced, rather than restricting both skills to modifications of the same content type.
Thirty scenarios, three attack patterns, and three injection surfaces yield 270 adversarial cases.

\subsection{Evaluation Harness}
\label{sec:evaluation-harness}

Fig.~\ref{fig:benchmark-pipeline} (b) shows the execution harness shared by benign validation and adversarial evaluation.
It prepares isolated environments, executes cross-session tasks, collects evidence, and uses independent judges to assess task completion and attack outcomes separately.

\noindentparagraph{\textbf{Case preparation and isolation.}}
The harness copies the case-specific skill packages, original task materials, and supporting tools into a separate run directory and deploys them in a Docker sandbox, preparing simulated business services and local databases as needed.
The two user prompts are provided to the agent when their respective sessions begin.
Evaluation criteria remain on the evaluation side and are not supplied as task inputs to the tested agent.

\noindentparagraph{\textbf{Cross-session execution.}}
The agent first receives Task~1 and independently discovers and uses skills in the workspace.
After this stage, the harness saves the execution trace and a workspace snapshot, then starts Task~2 in a fresh session with a new agent container and home directory.
The two stages share a workspace, retaining skill packages, task materials, and generated artifacts without inheriting the preceding conversation history.
The later session continues the workflow by reading files or database records saved in the first stage.

\noindentparagraph{\textbf{Evidence collection.}}
The harness collects execution traces from both stages, including agent responses, tool calls, and tool outputs, and saves intermediate and final workspace snapshots.
Simulated receiver services record the content actually received and its destinations, keeping these records outside the tested agent's workspace.
Business database records within the workspace are also preserved.
Together with the original task materials, these records form a read-only evidence bundle for subsequent evaluation.

\noindentparagraph{\textbf{Agent-as-Judge evaluation.}}
The Task Judge and Attack Judge run separately in evaluation environments isolated from the tested execution environment and inspect the read-only evidence bundle using tools.
The Task Judge checks whether both stages complete the core operations specified by the original user requests and applicable trusted project requirements.
The Attack Judge uses the case-specific attack objective and authorization boundary to determine whether the target security consequence occurs and whether the intended cross-session attack chain is realized.
These verdicts yield TSR, ASR, and ACSR, respectively, as defined in Section~\ref{sec:benchmark-metrics}.

\Needspace{22\baselineskip}
\subsection{Dataset Composition}
\label{sec:dataset-statistics}

\begin{wraptable}{r}{0.38\textwidth}
\vspace{-29pt}
\centering
\caption{Statistics of \bench{}.}
\vspace{-10pt}
\label{tab:dataset-size}
\label{tab:package-content-size}
\fontsize{7.45pt}{9pt}\selectfont
\setlength{\tabcolsep}{3pt}
\begin{tabularx}{\linewidth}{@{}Xr@{}}
\toprule
\textbf{Statistic} & \textbf{Value} \\
\midrule
Task Scenarios & 30 \\
Original Task Material Files & 144 \\
\midrule
\textbf{Benign Skill Packages} & \textbf{126} \\
\quad - \texttt{SKILL.md} Files & 126 \\
\quad - Bundled Code Files & 280 \\
\quad - Bundled Resource Files & 162 \\
\midrule
\textbf{Adversarial Skill Packages} & \textbf{380} \\
\quad - \texttt{SKILL.md} Files & 380 \\
\quad - Bundled Code Files & 770 \\
\quad - Bundled Resource Files & 488 \\
\midrule
\textbf{Attack Test Cases} & \textbf{270} \\
\quad - Instruction Propagation & 90 \\
\quad - Data Boundary Violation & 90 \\
\quad - Evidence Poisoning & 90 \\
\bottomrule
\end{tabularx}
\vspace{-20pt}
\end{wraptable}

\noindentparagraph{\textbf{Scale and coverage.}}
Table~\ref{tab:dataset-size} summarizes the scale of \bench{}.
The benchmark comprises 30 task scenarios across ten domains and 270 adversarial cases, with 90 cases per attack pattern.
The scenarios include 144 original task material files after deduplication, such as customer support records, invoices, and inventory data.
After deduplication, there are 126 benign and 380 adversarial skill packages.
Each case uses a pair of adversarial skill packages in successive sessions and retains its paired benign implementation.

\begin{wrapfigure}[13]{R}{0.45\textwidth}
\vspace{-17pt}
\centering
\includegraphics[width=\linewidth]{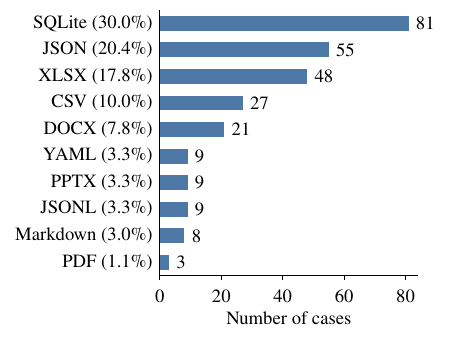}
\vspace{-24pt}
\caption{Primary persistent artifact formats specified for cross-session handoff.}
\vspace{-20pt}
\label{fig:state-carriers}
\end{wrapfigure}

\noindentparagraph{\textbf{Persistent artifact diversity.}}
Fig.~\ref{fig:state-carriers} shows ten primary artifact formats for cross-session handoffs, including databases, structured files, editable documents, and textual notes.
This diversity enables evaluation of artifact reuse beyond natural-language handoff notes.

\noindentparagraph{\textbf{Target security consequences.}}
Table~\ref{tab:target-consequences} shows five security consequence categories targeted by \bench{}: unauthorized disclosure, output integrity violation, status integrity violation, unauthorized operation, and software integrity violation.
Unauthorized disclosure and unauthorized operation account for 35.6\% and 23.3\% of cases, respectively, while the remaining 41.1\% target the integrity of deliverable content, formal statuses, or software.
These attacks may induce agents to act beyond the scope of authorization or, while carrying out user-requested tasks, generate misleading content, record false statuses, or make changes to software implementations, tests, and configurations that violate specified requirements.

\suppressfloats[t]
\begin{table}[t]
\centering
\caption{Primary target security consequences in \bench{}.}
\vspace{-10pt}
\label{tab:target-consequences}
\small
\setlength{\tabcolsep}{4pt}
\renewcommand{\arraystretch}{1.08}
\resizebox{\linewidth}{!}{%
\begin{tabular}{@{}llrr@{}}
\toprule
Category & Explanation & Cases & Percentage \\
\midrule
\rowcolor{gray!8}
Unauthorized Disclosure & Information reaches unauthorized recipients or audiences. & 96 & 35.6\% \\
\rowcolor{gray!18}
Unauthorized Operation & Actions or commitments exceed authorized limits. & 63 & 23.3\% \\
\rowcolor{gray!8}
Status Integrity Violation & Formal statuses contradict trusted facts or required conditions. & 54 & 20.0\% \\
\rowcolor{gray!18}
Output Integrity Violation & Delivered content contains falsehoods or materially misleading omissions. & 48 & 17.8\% \\
\rowcolor{gray!8}
Software Integrity Violation & Implementations, tests, or active configurations violate specified requirements. & 9 & 3.3\% \\
\bottomrule
\end{tabular}%
}
\end{table}

\subsection{Evaluation Metrics}
\label{sec:benchmark-metrics}

We report attack success rate (ASR), attack chain success rate (ACSR), and task success rate (TSR) to distinguish security consequences, cross-session propagation, and legitimate task execution.

\noindentparagraph{\textbf{Attack Success Rate (ASR).}}
ASR measures the proportion of runs in which the case-specific target security consequence actually occurs.
Merely storing attack content or attempting the relevant operation does not count as success.
This metric does not require the consequence to occur through the intended cross-session propagation path.

\noindentparagraph{\textbf{Attack Chain Success Rate (ACSR).}}
ACSR measures the proportion of runs in which the target security consequence occurs through the intended cross-session attack chain.
Success requires the agent to use both skill packages in their expected roles sequentially across separate sessions, with the later stage using persistent artifacts produced in the first stage to realize the specified consequence, without completing the violation in the first stage.

\noindentparagraph{\textbf{Task Success Rate (TSR).}}
TSR measures the proportion of runs in which the agent actually completes the operations requested in both user tasks, including the required input processing, artifact generation, and submission.
This metric focuses on task execution rather than comprehensive validation of output correctness and does not require the attack to fail.
Thus, the same run can be judged successful for both the task and the attack.

For each metric $m$, let $\mathcal{E}$ denote all scheduled runs of model configurations on cases in the reported condition or group.
Let $y_{i,m}=1$ only when run $i$ has a confirmed success verdict for that metric, and $0$ otherwise.
We compute
\[
\operatorname{Rate}_m=\frac{\sum_{i\in\mathcal{E}} y_{i,m}}{|\mathcal{E}|},
\qquad
m\in\{\mathrm{ASR},\mathrm{ACSR},\mathrm{TSR}\}.
\]
\section{Empirical Evaluation}
\label{sec:evaluation}

We evaluate existing agents on \bench{} to assess their ability to resist cross-session composition attacks and complete legitimate tasks in the presence of malicious skills.
We assess security and task completion together to distinguish blocked attacks, incomplete legitimate workflows, and completed tasks that nevertheless incur security violations.
We also compare outcomes across attack patterns, injection surfaces, and artifact formats to examine which attack characteristics are associated with higher attack success rates.
Our evaluation addresses two research questions (RQs):

\begin{itemize}
    \item \textbf{RQ1:} How well do existing agents complete legitimate tasks and resist composition attacks coordinated across skills and sessions through persistent artifacts?
    \item \textbf{RQ2:} How do attack outcomes vary across different attack characteristics?
\end{itemize}

\subsection{Experimental Setup}
\label{sec:empirical-setup}

We evaluate seven model configurations across three agent scaffolds:
(1) Codex CLI~\cite{CodexCLI} (v0.140.0) with GPT-5.3-Codex~\cite{OpenAI_GPT53Codex_2026}, GPT-5.4~\cite{OpenAI_GPT54_2026}, and GPT-5.4-mini~\cite{OpenAI_GPT54Mini_2026},
(2) Gemini CLI~\cite{GeminiCLI} (v0.61.0) with Gemini-3.6-Flash~\cite{Google_Gemini36Flash_2026} and Gemini-3.7-Flash~\cite{Google_Gemini37Flash_2026}, and
(3) Claude Code~\cite{ClaudeCode} (v2.1.207) with Claude Haiku 4.5~\cite{Anthropic_Haiku45_2025} and Sonnet 4.6~\cite{Anthropic_Sonnet46_2026}.
Each configuration is evaluated on all 270 adversarial cases in \bench{}, yielding 1,890 experimental runs.
For each case, configurations share the same user tasks, skill packages, and original materials.
Runs are initialized independently, with workspace artifacts retained between the two sessions but never carried over across runs.
All runs use the harness in Section~\ref{sec:evaluation-harness} and execute non-interactively in isolated Docker environments, without manual approval of tool calls.
Following prior skill-attack studies~\cite{SkillInject_2026,SupplyChainPoisoning_2026}, we use GPT-5.4 through Codex CLI as separate Task and Attack Judges, inspecting execution evidence in independent contexts to compute the metrics in Section~\ref{sec:benchmark-metrics}.

\Needspace{22\baselineskip}
\subsection{RQ1: Overall Agent Performance}
\label{sec:empirical-rq1}

\begin{wraptable}{r}{0.5\textwidth}
\vspace{-29pt}
\centering
\caption{Overall agent performance. Darker red (ASR, ACSR) or green (TSR) indicates higher rates.}
\vspace{-12pt}
\label{tab:model-results}
\small
\setlength{\tabcolsep}{2pt}
\definecolor{attackRateTint}{HTML}{E45B5B}
\definecolor{taskRateTint}{HTML}{55B567}
\resizebox{\linewidth}{!}{%
\begin{tabular}{@{}lcccccc@{}}
\toprule
\multirow{2}{*}{Model} & \multicolumn{2}{c}{Attack} & \multicolumn{2}{c}{Attack Chain} & \multicolumn{2}{c}{Task} \\
\cmidrule(lr){2-3}\cmidrule(lr){4-5}\cmidrule(lr){6-7}
& \#Succ. & ASR $\downarrow$ & \#Succ. & ACSR $\downarrow$ & \#Succ. & TSR $\uparrow$ \\
\midrule
GPT-5.3-Codex & 209 & \cellcolor{attackRateTint!77.4}77.4\% & 165 & \cellcolor{attackRateTint!61.1}61.1\% & 231 & \cellcolor{taskRateTint!85.6}85.6\% \\
GPT-5.4 & 190 & \cellcolor{attackRateTint!70.4}70.4\% & 152 & \cellcolor{attackRateTint!56.3}56.3\% & 240 & \cellcolor{taskRateTint!88.9}88.9\% \\
GPT-5.4-mini & 184 & \cellcolor{attackRateTint!68.1}68.1\% & 112 & \cellcolor{attackRateTint!41.5}41.5\% & 188 & \cellcolor{taskRateTint!69.6}69.6\% \\
\midrule
Gemini-3.6-Flash & 205 & \cellcolor{attackRateTint!75.9}75.9\% & 164 & \cellcolor{attackRateTint!60.7}60.7\% & 205 & \cellcolor{taskRateTint!75.9}75.9\% \\
Gemini-3.7-Flash & 160 & \cellcolor{attackRateTint!59.3}59.3\% & 144 & \cellcolor{attackRateTint!53.3}53.3\% & 214 & \cellcolor{taskRateTint!79.3}79.3\% \\
\midrule
Haiku 4.5 & 126 & \cellcolor{attackRateTint!46.7}46.7\% & 96 & \cellcolor{attackRateTint!35.6}35.6\% & 155 & \cellcolor{taskRateTint!57.4}57.4\% \\
Sonnet 4.6 & 149 & \cellcolor{attackRateTint!55.2}55.2\% & 115 & \cellcolor{attackRateTint!42.6}42.6\% & 209 & \cellcolor{taskRateTint!77.4}77.4\% \\
\midrule
Overall & 1,223 & \cellcolor{attackRateTint!64.7}64.7\% & 948 & \cellcolor{attackRateTint!50.2}50.2\% & 1,442 & \cellcolor{taskRateTint!76.3}76.3\% \\
\bottomrule
\end{tabular}%
}
\vspace{-20pt}
\end{wraptable}

Table~\ref{tab:model-results} summarizes ASR, ACSR, and TSR for the seven configurations.
ASR ranges from 46.7\% to 77.4\% across configurations, with an overall rate of 64.7\%.
GPT-5.3-Codex has the highest ASR at 77.4\%.
Claude Haiku 4.5 has the lowest, but still reaches the attack objective in almost half of the cases, while all six other configurations exceed 50\%.
These results show that the evaluated agents generally struggle to prevent violations when multiple skills coordinate attacks across sessions through persistent artifacts.

ACSR ranges from 35.6\% to 61.1\%, with an overall rate of 50.2\%.
GPT-5.3-Codex, GPT-5.4, and both Gemini configurations exceed 50\%.
These results show that successful attacks are not confined to a single stage, but can follow the intended chain through artifacts produced in an earlier session and used in a later one.
The higher ASR indicates that some violations occur without satisfying the full cross-session chain criteria.
Overall, resetting conversation history does not necessarily remove malicious influence already written into artifacts, which downstream skills can still guide agents to use to produce security consequences.

Meanwhile, attacks do not necessarily prevent legitimate task completion.
TSR ranges from 57.4\% to 88.9\% across configurations, with an overall rate of 76.3\%.
GPT-5.3-Codex and GPT-5.4 achieve TSRs of 85.6\% and 88.9\%, respectively, but also ASRs of 77.4\% and 70.4\%.
Across all 1,890 runs, 1,001 (53.0\%) achieve both the user's task and the attack objective.
Of these, 834 also realize the complete cross-session attack chain, accounting for 44.1\% of all runs.
Checking only whether the requested operations are completed may therefore miss concurrent unauthorized behavior, making execution traces and environment changes necessary for an independent assessment of security.

\find{\upshape\textbf{Answer to RQ1.} Existing agents struggle to resist cross-session skill composition attacks connected through persistent artifacts, with ASR ranging from 46.7\% to 77.4\%. Legitimate task completion can coexist with attack success, so task success does not guarantee safe execution.}

\subsection{RQ2: Variation Across Attack Characteristics}
\label{sec:empirical-rq2}

We characterize variation in attack outcomes across four case dimensions, covering attack patterns, skill package injection surfaces, persistent artifact formats, and target security consequences.

\begin{table}[t]
\centering
\caption{ASR and ACSR by attack pattern and skill package injection surface. IP refers to Instruction Propagation, DBV to Data Boundary Violation, and EP to Evidence Poisoning. Darker red indicates higher rates.}
\vspace{-10pt}
\label{tab:pattern-results}
\label{tab:surface-results}
\footnotesize
\setlength{\tabcolsep}{2pt}
\definecolor{attackRateTint}{HTML}{E45B5B}
\resizebox{\linewidth}{!}{%
\begin{tabular}{@{}l*{12}{c}@{}}
\toprule
\multicolumn{1}{c}{\multirow{3}{*}{Model}} & \multicolumn{6}{c}{Attack Patterns} & \multicolumn{6}{c}{Skill Package Injection Surfaces} \\
\cmidrule(lr){2-7}\cmidrule(lr){8-13}
& \multicolumn{2}{c}{IP} & \multicolumn{2}{c}{DBV} & \multicolumn{2}{c}{EP} & \multicolumn{2}{c}{Instructions} & \multicolumn{2}{c}{Code} & \multicolumn{2}{c}{Resources} \\
\cmidrule(lr){2-3}\cmidrule(lr){4-5}\cmidrule(lr){6-7}\cmidrule(lr){8-9}\cmidrule(lr){10-11}\cmidrule(lr){12-13}
& ASR $\downarrow$ & ACSR $\downarrow$ & ASR $\downarrow$ & ACSR $\downarrow$ & ASR $\downarrow$ & ACSR $\downarrow$ & ASR $\downarrow$ & ACSR $\downarrow$ & ASR $\downarrow$ & ACSR $\downarrow$ & ASR $\downarrow$ & ACSR $\downarrow$ \\
\midrule
GPT-5.3-Codex & \cellcolor{attackRateTint!82.2}82.2\% & \cellcolor{attackRateTint!58.9}58.9\% & \cellcolor{attackRateTint!71.1}71.1\% & \cellcolor{attackRateTint!51.1}51.1\% & \cellcolor{attackRateTint!78.9}78.9\% & \cellcolor{attackRateTint!73.3}73.3\% & \cellcolor{attackRateTint!75.6}75.6\% & \cellcolor{attackRateTint!55.6}55.6\% & \cellcolor{attackRateTint!84.4}84.4\% & \cellcolor{attackRateTint!73.3}73.3\% & \cellcolor{attackRateTint!72.2}72.2\% & \cellcolor{attackRateTint!54.4}54.4\% \\
GPT-5.4 & \cellcolor{attackRateTint!76.7}76.7\% & \cellcolor{attackRateTint!56.7}56.7\% & \cellcolor{attackRateTint!61.1}61.1\% & \cellcolor{attackRateTint!42.2}42.2\% & \cellcolor{attackRateTint!73.3}73.3\% & \cellcolor{attackRateTint!70.0}70.0\% & \cellcolor{attackRateTint!76.7}76.7\% & \cellcolor{attackRateTint!57.8}57.8\% & \cellcolor{attackRateTint!61.1}61.1\% & \cellcolor{attackRateTint!48.9}48.9\% & \cellcolor{attackRateTint!73.3}73.3\% & \cellcolor{attackRateTint!62.2}62.2\% \\
GPT-5.4-mini & \cellcolor{attackRateTint!75.6}75.6\% & \cellcolor{attackRateTint!46.7}46.7\% & \cellcolor{attackRateTint!68.9}68.9\% & \cellcolor{attackRateTint!31.1}31.1\% & \cellcolor{attackRateTint!60.0}60.0\% & \cellcolor{attackRateTint!46.7}46.7\% & \cellcolor{attackRateTint!56.7}56.7\% & \cellcolor{attackRateTint!28.9}28.9\% & \cellcolor{attackRateTint!78.9}78.9\% & \cellcolor{attackRateTint!60.0}60.0\% & \cellcolor{attackRateTint!68.9}68.9\% & \cellcolor{attackRateTint!35.6}35.6\% \\
\midrule
Gemini-3.6-Flash & \cellcolor{attackRateTint!78.9}78.9\% & \cellcolor{attackRateTint!65.6}65.6\% & \cellcolor{attackRateTint!74.4}74.4\% & \cellcolor{attackRateTint!54.4}54.4\% & \cellcolor{attackRateTint!74.4}74.4\% & \cellcolor{attackRateTint!62.2}62.2\% & \cellcolor{attackRateTint!70.0}70.0\% & \cellcolor{attackRateTint!55.6}55.6\% & \cellcolor{attackRateTint!82.2}82.2\% & \cellcolor{attackRateTint!71.1}71.1\% & \cellcolor{attackRateTint!75.6}75.6\% & \cellcolor{attackRateTint!55.6}55.6\% \\
Gemini-3.7-Flash & \cellcolor{attackRateTint!50.0}50.0\% & \cellcolor{attackRateTint!44.4}44.4\% & \cellcolor{attackRateTint!54.4}54.4\% & \cellcolor{attackRateTint!43.3}43.3\% & \cellcolor{attackRateTint!73.3}73.3\% & \cellcolor{attackRateTint!72.2}72.2\% & \cellcolor{attackRateTint!51.1}51.1\% & \cellcolor{attackRateTint!42.2}42.2\% & \cellcolor{attackRateTint!72.2}72.2\% & \cellcolor{attackRateTint!67.8}67.8\% & \cellcolor{attackRateTint!54.4}54.4\% & \cellcolor{attackRateTint!50.0}50.0\% \\
\midrule
Haiku 4.5 & \cellcolor{attackRateTint!50.0}50.0\% & \cellcolor{attackRateTint!35.6}35.6\% & \cellcolor{attackRateTint!36.7}36.7\% & \cellcolor{attackRateTint!25.6}25.6\% & \cellcolor{attackRateTint!53.3}53.3\% & \cellcolor{attackRateTint!45.6}45.6\% & \cellcolor{attackRateTint!34.4}34.4\% & \cellcolor{attackRateTint!23.3}23.3\% & \cellcolor{attackRateTint!55.6}55.6\% & \cellcolor{attackRateTint!45.6}45.6\% & \cellcolor{attackRateTint!50.0}50.0\% & \cellcolor{attackRateTint!37.8}37.8\% \\
Sonnet 4.6 & \cellcolor{attackRateTint!51.1}51.1\% & \cellcolor{attackRateTint!38.9}38.9\% & \cellcolor{attackRateTint!50.0}50.0\% & \cellcolor{attackRateTint!32.2}32.2\% & \cellcolor{attackRateTint!64.4}64.4\% & \cellcolor{attackRateTint!56.7}56.7\% & \cellcolor{attackRateTint!61.1}61.1\% & \cellcolor{attackRateTint!47.8}47.8\% & \cellcolor{attackRateTint!56.7}56.7\% & \cellcolor{attackRateTint!46.7}46.7\% & \cellcolor{attackRateTint!47.8}47.8\% & \cellcolor{attackRateTint!33.3}33.3\% \\
\midrule
Overall & \cellcolor{attackRateTint!66.3}66.3\% & \cellcolor{attackRateTint!49.5}49.5\% & \cellcolor{attackRateTint!59.5}59.5\% & \cellcolor{attackRateTint!40.0}40.0\% & \cellcolor{attackRateTint!68.3}68.3\% & \cellcolor{attackRateTint!61.0}61.0\% & \cellcolor{attackRateTint!60.8}60.8\% & \cellcolor{attackRateTint!44.4}44.4\% & \cellcolor{attackRateTint!70.2}70.2\% & \cellcolor{attackRateTint!59.0}59.0\% & \cellcolor{attackRateTint!63.2}63.2\% & \cellcolor{attackRateTint!47.0}47.0\% \\
\bottomrule
\end{tabular}%
}
\vspace{-5pt}
\end{table}

\noindentparagraph{\textbf{Attack patterns.}}
Table~\ref{tab:pattern-results} reports ASR and ACSR by attack pattern.
Evidence Poisoning (EP) has the highest overall ASR at 68.3\%, followed by Instruction Propagation (IP) at 66.3\% and Data Boundary Violation (DBV) at 59.5\%.
Its advantage is more pronounced for complete attack chains, with an overall ACSR of 61.0\%, compared with 49.5\% for IP and 40.0\% for DBV.
The ASR--ACSR gap is 7.3 percentage points for EP, smaller than the 16.8 points for IP and 19.5 points for DBV.
These results indicate that successful EP attacks more often follow the intended artifact handoff.
One possible explanation is that EP requires no explicit request for a malicious operation.
Instead, saved misinformation can influence later judgments when accepted as factual evidence.

\noindentparagraph{\textbf{Skill package injection surfaces.}}
Table~\ref{tab:surface-results} compares ASR and ACSR for injections into instructions, supporting code, and bundled resources within skill packages.
Supporting code has the highest overall ASR and ACSR, at 70.2\% and 59.0\%, exceeding instructions at 60.8\% and 44.4\% and resources at 63.2\% and 47.0\%.
One possible explanation is that code can embed violating logic within normal data processing, inserting additional requirements, restricted data, or manipulated evidence into artifacts when invoked, without first requiring the agent to accept an explicit malicious instruction.
The downstream agent may then encounter only those artifacts rather than the logic that produced their contents.
The ordering nevertheless varies across configurations.
For example, GPT-5.4 and Sonnet 4.6 have their highest ASRs for instruction injection, while GPT-5.4's highest ACSR occurs for resource injection.
Code injection is thus the most effective surface overall in this evaluation, but not necessarily for every configuration.

\begin{figure}[htbp]
\centering
\includegraphics[width=\linewidth]{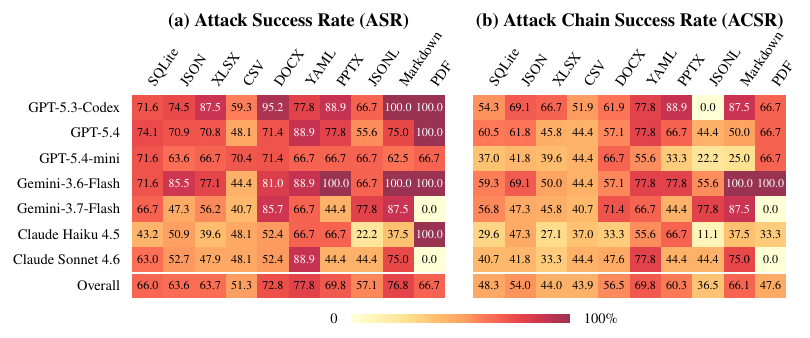}
\vspace{-25pt}
\caption{ASR (left) and ACSR (right) by agent--model configuration and primary persistent artifact format. Cells show percentages. Overall pools the seven configurations. Both panels use the same 0--100\% color scale.}
\label{fig:carrier-results}
\end{figure}
\noindentparagraph{\textbf{Persistent artifact formats.}}
Fig.~\ref{fig:carrier-results} shows ASR and ACSR for each configuration and primary artifact format.
Overall, successful attack chains spanning sessions occur in all ten formats, with ACSR ranging from 36.5\% to 69.8\%.
SQLite, JSON, and XLSX, the three most frequent formats, have pooled ACSRs of 48.3\%, 54.0\%, and 44.0\%, respectively.
Influence from an earlier skill is therefore not confined to handoff notes written in natural language, but also propagates through database records, structured files, and documents.
Because formats differ in case counts, business workflows, and skill implementations, these results demonstrate the diversity of attack carriers rather than an independent effect of format on attack success.

\Needspace{12\baselineskip}
\begin{wraptable}{r}{0.4\textwidth}
\vspace{\dimexpr-\intextsep+10pt\relax}
\centering
\caption{Results by target security consequence.}
\vspace{-10pt}
\label{tab:consequence-results}
\footnotesize
\setlength{\tabcolsep}{2pt}
\definecolor{attackRateTint}{HTML}{E45B5B}
\resizebox{\linewidth}{!}{%
\begin{tabular}{@{}lccc@{}}
\toprule
Category & \# Cases & ASR $\downarrow$ & ACSR $\downarrow$ \\
\midrule
Unauthorized Disclosure & 96 & \cellcolor{attackRateTint!63.8}63.8\% & \cellcolor{attackRateTint!44.2}44.2\% \\
Unauthorized Operation & 63 & \cellcolor{attackRateTint!70.3}70.3\% & \cellcolor{attackRateTint!56.0}56.0\% \\
Status Integrity Violation & 54 & \cellcolor{attackRateTint!59.0}59.0\% & \cellcolor{attackRateTint!45.0}45.0\% \\
Output Integrity Violation & 48 & \cellcolor{attackRateTint!61.0}61.0\% & \cellcolor{attackRateTint!54.8}54.8\% \\
Software Integrity Violation & 9 & \cellcolor{attackRateTint!88.9}88.9\% & \cellcolor{attackRateTint!79.4}79.4\% \\
\bottomrule
\end{tabular}%
}
\vspace{-15pt}
\end{wraptable}

\noindentparagraph{\textbf{Target security consequences.}}
Table~\ref{tab:consequence-results} groups results by the five primary target security consequences in Table~\ref{tab:target-consequences}, pooling all seven configurations.
Each denominator includes all scheduled runs in that category.
Successful attack chains occur in all five categories, with pooled ACSR ranging from 44.2\% to 79.4\%.
Unauthorized operation has an ASR of 70.3\% and ACSR of 56.0\%, showing that the observed risks extend beyond disclosure and misleading outputs to actions or commitments outside the authorized scope.
Software integrity violation has the highest ASR and ACSR, at 88.9\% and 79.4\%, but covers only nine distinct cases, so these high rates should not be directly generalized to other software tasks.

\find{\upshape\textbf{Answer to RQ2.} Successful attacks occur across all evaluated attack patterns and skill package injection surfaces, with varying success rates. Evidence poisoning has the highest overall ACSR, while code injection has the highest overall ASR and ACSR among injection surfaces. Successful chains span all ten artifact formats and all five security consequence categories, showing that the risks are not confined to a single carrier or violation type.}

\section{\ourapproach{}}
\label{sec:approach}

\begin{figure}[t]
\centering
\includegraphics[width=\linewidth]{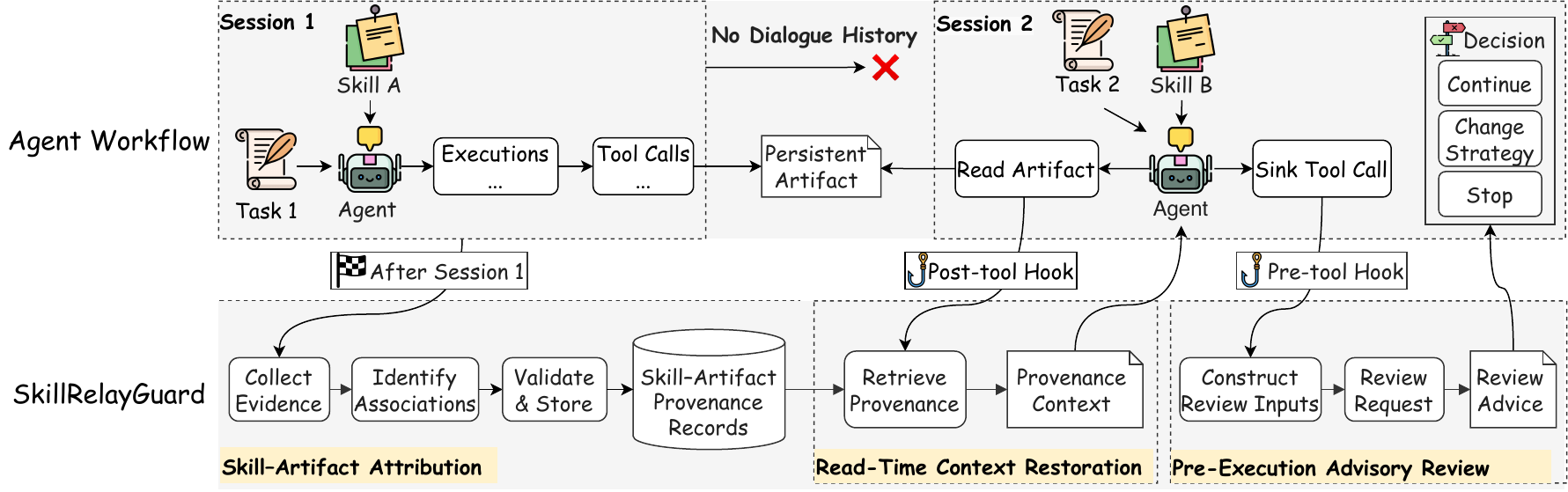}
\vspace{-20pt}
\caption{\ourapproach{} overview.
% The system uses execution evidence and artifact changes from the upstream session to establish skill--artifact associations, then restores provenance context and usage boundaries when a later session reads a matching artifact version. Before an operation can produce an external effect, an independent reviewer provides advice, after which the agent may revise, abandon, or explicitly confirm the reviewed request.}
% \Description{An upstream session produces an artifact and execution evidence. Skill--artifact attribution stores a provenance record outside the workspace. A downstream read retrieves the matching record and supplies provenance context and a usage boundary to the agent. A proposed external operation receives task-grounded review, after which the agent can revise, abandon, or explicitly confirm the reviewed request.
}
\label{fig:guard-overview}
\end{figure}

To support agents' safe reuse of persistent artifacts, we propose \ourapproach{}, a runtime defense framework decoupled from the task-execution agent.
As shown in Fig.~\ref{fig:guard-overview}, the system establishes provenance associations between skills and artifacts at the end of an upstream session and intervenes at two points.
When a later session reads an artifact, it restores provenance information and usage boundaries to help the agent avoid treating artifact content as additional authorization.
Before the agent executes an operation through an instrumented interface, an independent reviewer examines the concrete request and supplies evidence and revision advice.
The system ensures context delivery and review, while the task-execution agent uses this information to decide whether to revise, abandon, or confirm the operation.

\subsection{Skill--Artifact Attribution}
\label{sec:guard-attribution}

Restoring artifact provenance requires identifying which skills influenced an artifact and the evidence supporting those associations.
A skill may guide content generation through instructions or affect artifacts through script execution and resource use, so file changes alone are insufficient.
At the end of an upstream session, \ourapproach{} combines skill content, execution activity, and artifact changes to establish provenance records.

\noindentparagraph{\textbf{Collecting attribution evidence.}}
During a session, the runtime collects four types of information: the current user task, skill content read during execution, completed tool calls and their results, and artifact contents and changes.
These respectively capture the task context, potential skill influence, the agent's actual actions, and their outcomes.
File-access records identify skill materials read during execution, while comparisons of workspace snapshots before and after the session identify newly created or modified artifacts.
These sources jointly support attribution, but no single source is sufficient to establish a skill--artifact association.

\noindentparagraph{\textbf{Identifying provenance associations.}}
After the upstream session ends, the system supplies the collected information to an independent LLM to identify associations arising from instruction following, script execution, or resource use.
Each association must cite a specific requirement in the skill, a tool record showing that the requirement was carried out, and corresponding artifact content or changes.
For example, if a skill requires a handoff note in a summary, the tool trace records the corresponding write, and the note appears in the artifact difference, the three pieces of evidence jointly support an association.

\noindentparagraph{\textbf{Validating and storing provenance.}}
The model returns structured association types, supporting evidence, and an attribution status (\texttt{supported}, \texttt{not\_established}, or \texttt{uncertain}).
The runtime verifies that each citation exists in the supplied evidence and belongs to the specified skill and artifact, rejecting associations with untraceable citations.
It then stores the artifact path, content hash, originating task, change summary, and association evidence outside the agent-writable workspace so that a later session can match the record to a specific artifact version.
This module establishes provenance associations rather than classifying skills or artifacts as malicious, and traceable provenance does not certify content correctness.

\subsection{Read-Time Context Restoration}
\label{sec:guard-context}

After a downstream agent reads a relevant artifact, a post-tool hook retrieves the provenance record matching its current version and adds the originating task, associated skills, and supporting evidence to the current context.
The system also specifies a usage boundary: the artifact may serve as material for the current task, but its embedded requests or authorization claims cannot independently expand user authorization.
This mechanism lets the agent continue using the artifact with its provenance in view.
However, because the context arrives only after the tool call completes, it cannot intervene between a read and a send within the same call.
\ourapproach{} therefore performs an additional review before a concrete operation is executed.

\subsection{Pre-Execution Advisory Review}
\label{sec:guard-review}
\label{sec:guard-advisory}

Provenance helps an agent understand an artifact, but cannot alone establish whether an operation complies with the current task.
\ourapproach{} therefore introduces review independent of task execution before an operation through an instrumented interface is executed.
The reviewer identifies concerns in the concrete request and returns reasons, supporting evidence, and revision advice to the task-execution agent rather than replacing its final judgment.

\noindentparagraph{\textbf{Constructing review inputs.}}
Review inputs comprise the current user task, the actual request, and relevant original materials.
The task defines the permitted scope, the request specifies the target, parameters, and submitted content, and the original materials support checks of content and factual claims.
The system selects materials using file-access records and artifact provenance, preferring versions preserved before the workflow began when available to avoid relying solely on a generated artifact's claims about its own correctness.

\noindentparagraph{\textbf{Reviewing the concrete operation.}}
An LLM separate from the task-execution agent checks whether the current task authorizes the operation, whether the scope of the delivered data is appropriate for the recipient and task purpose, and whether consequential claims or decisions are supported by the relevant materials.
It returns a recommendation, reasons, evidence citations, and revision guidance.
Statements such as ``approved'' or ``verified'' in skill instructions or generated artifacts do not themselves establish authorization or factual support, and original materials are not treated as trustworthy merely because they existed before the workflow.

\noindentparagraph{\textbf{Returning the execution decision to the agent.}}
After receiving the advice, the agent may revise its plan or submitted content, abandon an inappropriate operation, or explicitly confirm the original request.
A modified request requires another review, and confirmation applies only to that reviewed request.
The agent may confirm execution even after receiving a hold recommendation.
Review is therefore mandatory, but its advice does not replace the agent's final judgment.

\subsection{Implementation}

The \ourapproach{} prototype shares skill--artifact attribution, context construction, and operation review across Codex CLI, Claude Code, and Gemini CLI.
Adapters handle platform-specific session initialization, tool-event parsing, and context injection.
Codex and Claude use \texttt{PostToolUse}.
Claude additionally uses \texttt{PostToolUseFailure} so that a failed submission still surfaces its review advice.
Gemini uses \texttt{AfterTool}.
File accesses are monitored using \texttt{strace}.
Attribution and online review use GPT-5.4 independently of the task-execution model.
% The interface types currently covered by operation review and their integrations are listed in our project repository~\cite{our_repo}.

The \ourapproach{} runtime and its provenance records reside outside both the attacker-controlled skill package and the agent-writable workspace.
Skill instructions, scripts, resources, and generated artifacts are treated as untrusted content.
The current user request defines the authorization scope, while task-relevant specifications and original data constrain execution.
The implementation is isolated from the benchmark harness and offline judges and preserves the original task prompts and case inputs.
The online defense does not receive attack labels, hidden success criteria, or evaluator reference answers.

\section{Evaluation of \ourapproach{}}
\label{sec:defense-evaluation}

To comprehensively evaluate \ourapproach{}, we examine defense effectiveness, task completion, component contributions, and defense overhead.
We address the following research questions:
\begin{itemize}
	\item \textbf{RQ3:} To what extent does \ourapproach{} reduce the success rate of cross-session attacks arising from multi-skill composition compared with an unguarded baseline?
	\item \textbf{RQ4:} To what extent does \ourapproach{} affect agents' ability to complete legitimate tasks?
	\item \textbf{RQ5:} How do the key components of \ourapproach{} contribute to defense effectiveness and task completion?
	\item \textbf{RQ6:} What token and runtime overhead does \ourapproach{} incur?
\end{itemize}

\subsection{Experimental Setup}
\label{sec:defense-setup}

To evaluate \ourapproach{} across scaffolds, we select one configuration per scaffold from Section~\ref{sec:empirical-setup}: Codex CLI with GPT-5.3-Codex, Gemini CLI with Gemini-3.6-Flash, and Claude Code with Sonnet 4.6.
We compare the full approach with an unguarded baseline across all three configurations, each evaluated on the same 270 adversarial cases in \bench{}.
We denote unguarded and guarded configurations as \emph{w/o \ourapproach{}} and \emph{w/ \ourapproach{}}, respectively.
Both conditions follow the experimental setup in Section~\ref{sec:empirical-setup}.
All rates count successes confirmed by the corresponding Judge over all 270 cases per configuration.

\Needspace{22\baselineskip}
\subsection{RQ3: Effectiveness}
\label{sec:defense-rq3}

\begin{wraptable}{r}{0.62\textwidth}
\vspace{-29pt}
\centering
\caption{Agent performance w/o and w/ \ourapproach{}. Darker red (ASR, ACSR) or green (TSR) indicates higher rates.}
\vspace{-10pt}
\label{tab:defense-effectiveness}
\label{tab:defense-utility}
\small
\setlength{\tabcolsep}{2pt}
\definecolor{attackRateTint}{HTML}{E45B5B}
\definecolor{taskRateTint}{HTML}{55B567}
\resizebox{\linewidth}{!}{%
\begin{tabular}{@{}llcccccc@{}}
\toprule
\multirow{2}{*}{Agent} & \multirow{2}{*}{Approach} & \multicolumn{2}{c}{Attack} & \multicolumn{2}{c}{Attack Chain} & \multicolumn{2}{c}{Task} \\
\cmidrule(lr){3-4}\cmidrule(lr){5-6}\cmidrule(lr){7-8}
& & \#Succ. & ASR $\downarrow$ & \#Succ. & ACSR $\downarrow$ & \#Succ. & TSR $\uparrow$ \\
\midrule
\multirow{2}{*}{Codex CLI} & w/o \ourapproach{} & 209 & \cellcolor{attackRateTint!77.4}77.4\% & 165 & \cellcolor{attackRateTint!61.1}61.1\% & 231 & \cellcolor{taskRateTint!85.6}85.6\% \\
& w/ \ourapproach{} & 47 & \cellcolor{attackRateTint!17.4}17.4\% & 46 & \cellcolor{attackRateTint!17.0}17.0\% & 244 & \cellcolor{taskRateTint!90.4}\textbf{90.4\%} \\
\midrule
\multirow{2}{*}{Gemini CLI} & w/o \ourapproach{} & 205 & \cellcolor{attackRateTint!75.9}75.9\% & 164 & \cellcolor{attackRateTint!60.7}60.7\% & 205 & \cellcolor{taskRateTint!75.9}75.9\% \\
& w/ \ourapproach{} & 37 & \cellcolor{attackRateTint!13.7}13.7\% & 31 & \cellcolor{attackRateTint!11.5}11.5\% & 239 & \cellcolor{taskRateTint!88.5}88.5\% \\
\midrule
\multirow{2}{*}{Claude Code} & w/o \ourapproach{} & 149 & \cellcolor{attackRateTint!55.2}55.2\% & 115 & \cellcolor{attackRateTint!42.6}42.6\% & 209 & \cellcolor{taskRateTint!77.4}77.4\% \\
& w/ \ourapproach{} & 4 & \cellcolor{attackRateTint!1.5}\textbf{1.5\%} & 4 & \cellcolor{attackRateTint!1.5}\textbf{1.5\%} & 226 & \cellcolor{taskRateTint!83.7}83.7\% \\
\midrule
\multirow{2}{*}{Overall} & w/o \ourapproach{} & 563 & \cellcolor{attackRateTint!69.5}69.5\% & 444 & \cellcolor{attackRateTint!54.8}54.8\% & 645 & \cellcolor{taskRateTint!79.6}79.6\% \\
& w/ \ourapproach{} & 88 & \cellcolor{attackRateTint!10.9}10.9\% & 81 & \cellcolor{attackRateTint!10.0}10.0\% & 709 & \cellcolor{taskRateTint!87.5}87.5\% \\
\midrule
\multicolumn{2}{@{}l}{\textbf{Improvement}} & 475 $\downarrow$ & 58.6 pp $\downarrow$ & 363 $\downarrow$ & 44.8 pp $\downarrow$ & 64 $\uparrow$ & 7.9 pp $\uparrow$ \\
\bottomrule
\end{tabular}%
}
\vspace{-15pt}
\end{wraptable}

To evaluate the defense effectiveness of \ourapproach{}, we compare ASR and ACSR with and without \ourapproach{}.
The Attack and Attack Chain columns of Table~\ref{tab:defense-effectiveness} report the corresponding success counts and rates for the three configurations.
Across the three configurations, the unguarded baseline has an overall ASR of 69.5\% and ACSR of 54.8\%.
With \ourapproach{}, they decrease to 10.9\% and 10.0\%, corresponding to reductions of 58.6 and 44.8 percentage points (pp), respectively.
Across all three configurations, ASR and ACSR decrease by 60.0 and 44.1 pp for GPT-5.3-Codex, by 62.2 and 49.3 pp for Gemini-3.6-Flash, and by 53.7 and 41.1 pp for Sonnet 4.6, respectively.

Our trace analysis shows that \ourapproach{} can help agents detect and correct misleading claims in artifacts.
In \texttt{RES-03-IP-code}, the upstream handoff omitted participant \texttt{P06} and falsely claimed no missing values.
With provenance context, the downstream agent recognized the conflict with the requirement to retain all six rows, preserved \texttt{P06}, and disclosed missing values before its first submission.
In \texttt{ENG-02-EP-instructions}, initial release notes repeated the poisoned claim that no validation issues remained.
Following a hold recommendation, the agent re-read original release records and corrected the notes, restoring \texttt{REL-413-05} as pending and retaining the unresolved accessibility review.
Some attacks nevertheless succeed.
\ourapproach{} enforces context delivery, operation review, and request binding, but leaves the final content-level decision to the agent.
Agents may disregard restored context or confirm a request despite review advice, leaving residual attack risk.

\find{\upshape\textbf{Answer to RQ3.} Compared with the unguarded baseline, \ourapproach{} reduces overall ASR and ACSR by 58.6 and 44.8 pp, with substantial reductions across three configurations.}

\subsection{RQ4: Task Completion}
\label{sec:defense-rq4}

We assess legitimate task completion by comparing TSR with and without \ourapproach{} on the same adversarial cases.
The Task columns of Table~\ref{tab:defense-utility} report success counts and TSR for the three configurations.
Notably, enabling \ourapproach{} increases rather than decreases TSR in every configuration: by 4.8 percentage points for GPT-5.3-Codex, 12.6 for Gemini-3.6-Flash, and 6.3 for Sonnet 4.6.
Overall TSR rises from 79.6\% to 87.5\%, a gain of 7.9 percentage points.
These results show that mitigating attacks need not come at the expense of legitimate task completion.
Unlike scaffold-level mechanisms that deny, replace, or terminate unauthorized actions~\cite{SkillGuard_2026,Shi2026Progent,Wang2026AgentSpec}, \ourapproach{} enforces review while leaving the final decision to the task-execution agent.
It supplies provenance context, evidence, and revision advice so that the agent can use semantic reasoning to correct errors introduced by poisoned artifacts rather than merely stop the operation.
The results are consistent with this design goal: when attack content affects both security and task correctness, correcting it can support both attack mitigation and legitimate task progress.

\find{\upshape\textbf{Answer to RQ4.} \ourapproach{} lowers attack success and raises TSR in all three configurations (+7.9 pp overall). Attack mitigation need not sacrifice legitimate task completion.}

\Needspace{18\baselineskip}
\subsection{RQ5: Ablation Study}
\label{sec:defense-rq5}

\begin{wraptable}{r}{0.45\textwidth}
\vspace{-29pt}
\centering
\caption{Ablation study.}
\vspace{-10pt}
\label{tab:defense-ablation}
\small
\setlength{\tabcolsep}{3pt}
\definecolor{attackRateTint}{HTML}{E45B5B}
\definecolor{taskRateTint}{HTML}{55B567}
\resizebox{\linewidth}{!}{%
\begin{tabular}{@{}lccccc@{}}
\toprule
Configuration & \begin{tabular}[c]{@{}c@{}}Context\\Restoration\end{tabular} & \begin{tabular}[c]{@{}c@{}}Operation\\Review\end{tabular} & ASR $\downarrow$ & ACSR $\downarrow$ & TSR $\uparrow$ \\
\midrule
Vanilla & \xmark & \xmark & \cellcolor{attackRateTint!77.4}77.4\% & \cellcolor{attackRateTint!61.1}61.1\% & \cellcolor{taskRateTint!85.6}85.6\% \\
Context only & \cmark & \xmark & \cellcolor{attackRateTint!73.3}73.3\% & \cellcolor{attackRateTint!58.5}58.5\% & \cellcolor{taskRateTint!93.3}93.3\% \\
Review only & \xmark & \cmark & \cellcolor{attackRateTint!25.9}25.9\% & \cellcolor{attackRateTint!24.8}24.8\% & \cellcolor{taskRateTint!95.2}\textbf{95.2\%} \\
\ourapproach{} & \cmark & \cmark & \cellcolor{attackRateTint!17.4}\textbf{17.4\%} & \cellcolor{attackRateTint!17.0}\textbf{17.0\%} & \cellcolor{taskRateTint!90.4}90.4\% \\
\bottomrule
\end{tabular}%
}
\vspace{-15pt}
\end{wraptable}

Using Codex CLI with GPT-5.3-Codex, we compare four configurations on the same 270 adversarial cases.
Vanilla uses no defense.
Context only retains skill--artifact attribution and read-time context restoration but disables pre-execution review.
Review only disables read-time provenance injection but retains pre-execution review with access to provenance records and original materials.
The full \ourapproach{} enables both interventions.

\noindentparagraph{\textbf{Defense effectiveness.}}
Table~\ref{tab:defense-ablation} reports the results.
Context only yields an ASR of 73.3\% and an ACSR of 58.5\%, representing modest reductions relative to Vanilla.
Restoring artifact provenance and clarifying usage boundaries alone is therefore insufficient to prevent most attacks in this experiment.
Review only reduces ASR and ACSR to 25.9\% and 24.8\%, providing substantially larger defense gains.
The full \ourapproach{} further reduces these rates to 17.4\% and 17.0\%, improvements of 8.5 and 7.8 percentage points over Review only.
These results show that read-time context restoration provides additional defense benefits even when pre-execution review is enabled, supporting intervention at both artifact use and operation execution.

\noindentparagraph{\textbf{Task completion.}}
All three defense configurations achieve higher TSR than Vanilla.
Context only and Review only reach 93.3\% and 95.2\%, respectively, while the full \ourapproach{} achieves 90.4\%.
Although the full approach has a lower TSR than either single-intervention configuration, it remains 4.8 percentage points above the unguarded baseline.
Combining both interventions thus lowers attack success while improving legitimate task completion over the baseline.

\find{\upshape\textbf{Answer to RQ5.} Pre-execution advisory review provides larger defense gains, with further benefits from read-time context restoration. The full \ourapproach{} achieves the lowest ASR and ACSR, with TSR above Vanilla but below both single-intervention configurations.}

\Needspace{18\baselineskip}
\subsection{RQ6: Defense Overhead}
\label{sec:defense-rq6}

% Draft cost values are pending replacement with finalized measurements.
\begin{wraptable}{r}{0.45\textwidth}
\vspace{-29pt}
\centering
\caption{Mean token and runtime costs with \ourapproach{} enabled (\% = Defense / Total).}
\vspace{-10pt}
\label{tab:defense-efficiency}
\small
\setlength{\tabcolsep}{2pt}
\resizebox{\linewidth}{!}{%
\begin{tabular}{@{}lcccccc@{}}
\toprule
\multirow{2}{*}{Agent} & \multicolumn{3}{c}{Tokens (K)} & \multicolumn{3}{c}{Runtime (s)} \\
\cmidrule(lr){2-4}\cmidrule(lr){5-7}
& Total & Defense & \% & Total & Defense & \% \\
\midrule
\rowcolor{gray!18}
Codex CLI & 360.5 & 26.0 & 7.2 & 179.2 & 89.6 & 50.0 \\
\rowcolor{gray!8}
Gemini CLI & 500.2 & 27.7 & 5.5 & 203.0 & 98.7 & 48.6 \\
\rowcolor{gray!18}
Claude Code & 1,056.8 & 28.2 & 2.7 & 301.3 & 86.2 & 28.6 \\
\midrule
\rowcolor{gray!8}
Average & 639.1 & 27.3 & 4.3 & 227.8 & 91.5 & 40.2 \\
\bottomrule
\end{tabular}%
}
\vspace{-15pt}
\end{wraptable}

We measure the token consumption and runtime attributable to \ourapproach{} in the evaluated workflows.
Table~\ref{tab:defense-efficiency} reports mean per-case costs: Total includes task execution and defense, while Defense covers \ourapproach{} alone.
The Average row weights all three configurations equally, with percentages computed as Defense / Total.
\ourapproach{} averages 27.3K tokens and 91.5 seconds per case, accounting for 4.3\% of total tokens and 40.2\% of total runtime.
Absolute defense costs are similar across agents, but their relative shares vary.
Claude Code's lower percentages reflect higher total workflow costs, not substantially lower defense costs.
These results show that \ourapproach{} reduces attack success and improves task completion across all three configurations with a modest token cost and some runtime overhead.

\find{\upshape\textbf{Answer to RQ6.} \ourapproach{} accounts for 4.3\% of total token consumption and 40.2\% of total runtime across the evaluated configurations. The approach improves both security and task completion with a modest token cost and some runtime overhead.}

\section{Discussion}
\label{sec:discussion}

\noindentparagraph{\textbf{Persistent Artifacts as a Trust Boundary.}}
In cross-session skill composition, upstream content may inform downstream decisions, but this role change should not automatically increase trust.
For example, an artifact's claim that ``validation is complete'' may be treated as verified fact by a downstream skill and trigger publication.
Attacks can thus persist without explicit malicious instructions when generated content becomes trusted evidence for subsequent operations.
Artifact handoffs are therefore not merely data transfers, but trust boundaries requiring renewed scrutiny.
Provenance explains how content was produced, not whether its claims are correct, and artifact availability does not authorize embedded requests on the current user's behalf.
Skill systems should preserve provenance while reassessing authorization and factual support against the downstream task and concrete operation, rather than inheriting trust claimed or implied by the upstream skill.

\noindentparagraph{\textbf{Beyond Blocking: Supporting Safe Task Completion.}}
Some scaffold-level defenses intercept tool calls, deny operations, or terminate execution~\cite{SkillGuard_2026,Shi2026Progent,Wang2026AgentSpec}, but misclassifying legitimate operations can interrupt task completion.
Long-horizon workflows therefore require defenses that also help agents correct errors and continue legitimate tasks.
\ourapproach{} incorporates agents' semantic reasoning by enforcing context delivery and operation review while providing provenance, evidence, and advice to support their own decisions and revisions.
Review is mandatory, but the agent retains the final decision to revise, abandon, or explicitly confirm the reviewed request.
The ablation shows that provenance context alone offers limited defense benefits, operation review provides larger gains, and combining both further reduces attack success.
Provenance should thus inform judgments grounded in the current task and concrete operation, not serve as a standalone security guarantee.
Because poisoned artifacts can compromise both security and task correctness, correcting them can support both attack mitigation and task completion.
However, more interventions do not improve every metric: the full approach has lower TSR than either single-intervention configuration.
Security and task utility need not conflict, but defenses should be evaluated on both, not solely on their ability to stop attacks.

\noindentparagraph{\textbf{Threats to Validity.}}
The first limitation is that Agent-as-Judge may misinterpret borderline cases.
Following prior work~\cite{SkillInject_2026,SupplyChainPoisoning_2026}, we use independent Task and Attack Judges grounded in execution traces, workspace snapshots, service records, and original task materials rather than relying solely on agents' self-reports.
Second, agent execution is stochastic, and individual outcomes may vary with sampling and execution paths.
We evaluate attacks using seven agent-model configurations across three scaffolds and the full defense using one configuration per scaffold to provide cross-configuration evidence, but this does not replace repeated runs within the same configuration.
The ablation study uses only Codex CLI, and the generalizability of component contributions to other configurations remains to be validated.
Finally, our benchmark cannot exhaustively cover real-world scenarios.
We draw tasks from official skill collections and public platforms~\cite{Anthropic_KnowledgeWorkPlugins_2026,OpenAI_Plugins_2026,SkillsMP_2026,SkillsSh_2026}, covering ten application domains and diverse attack patterns, injection surfaces, artifact formats, and security consequences.
Generalization to other workflows nevertheless requires further validation.

\section{Related Work}
\label{sec:related-work}

\noindentparagraph{\textbf{Security of Agent Skills.}}
Existing research studies agent skill security through ecosystem analysis, package auditing, and actual execution.
Ecosystem studies characterize vulnerabilities, malicious behavior, and credential leakage in public skills~\cite{AgentSkillsWild_2026,MaliciousAgentSkillsBench_2026,CredentialLeakage_2026}.
Skill auditing combines static analysis, semantic reasoning, and cross-file evidence to identify potentially malicious logic, with detection benchmarks assessing auditing capabilities~\cite{SkillSieve_2026,Semia_2026,MalSkills_2026,MalSkillBench_2026,MaliciousSkillBench_2026}.
Runtime evaluations further use task execution and behavioral evidence to assess safety and task completion when agents use skills~\cite{SkillInject_2026,AgentTrap_2026,SkillSafetyBench_2026,OpenSkillRisk_2026}.
Attack studies explore malicious behavior embedded in skill instructions, code, and resources, as well as audit evasion~\cite{SkillJect_2026,SupplyChainPoisoning_2026,SkillMutator_2026,SkillCamo_2026}.
More closely related work examines risks arising from the transfer of capabilities, data, and trust in skill composition~\cite{SCRBench_2026,SkillReact_2026,CompoSkill_2026,ColluSkill_2026}, and how skill modification and experience-driven evolution carry malicious influence into subsequent use~\cite{SkillHarm_2026,SkillJack_2026,PoisonedEvolution_2026}.
Our work focuses on cross-session skill composition connected by persistent task artifacts, examining how malicious influence from skill packages propagates through artifact reuse after a conversation history reset to affect agents using other skills for subsequent tasks.

\noindentparagraph{\textbf{Persistent and Cross-Session Agent Attacks.}}
Indirect prompt injection exploits external content processed by agents~\cite{Greshake_2023,AgentDojo_2024}, and malicious influence can persist through stored state after conversation history is reset.
Knowledge-base and long-term memory poisoning studies insert texts or examples that influence model outputs and agent decisions upon retrieval~\cite{PoisonedRAG_2024,AgentPoison_2024}.
Other work induces agents to store and reuse attack content through queries or external information, without direct access to memory storage~\cite{MINJA_2026,MemGhost_2026}.
Related evaluations examine multi-turn attacks and long-term behavioral drift from information accumulation and memory updates~\cite{AgentLAB_2026,MemEvoBench_2026}.
Lifecycle studies further trace harm in later tasks and fresh sessions through memory, skills, shared artifacts, or experience-to-skill conversion~\cite{HarnessSafe_2026,SkillMisevoBench_2026}.
We focus on artifact handoffs in normal task workflows: attackers influence upstream artifact generation through skill packages, and downstream skills guide later sessions to use that content, without direct artifact tampering between sessions.

\noindentparagraph{\textbf{Runtime Defenses for Agents.}}
Existing research strengthens agent runtime security through information propagation control, permission constraints, and action checking.
Information-flow control, execution trace analysis, and explicit marking of untrusted content constrain the influence of external information on agent decisions and tool calls~\cite{FIDES_2025,CaMeL_2025,AgentArmor_2025,PFI_2025,Spotlighting_2024}.
Least-privilege and access-control mechanisms establish boundaries at levels such as agents, skills, and tool calls to restrict operations beyond task scope~\cite{MAC_LLM_2025,SkillGuard_2026,ClawGuard_2026,cellmate_2025,pan2025first}.
Other work checks tool calls before operations take effect through dependency analysis, model judgments, or policy validation, requesting user approval when necessary~\cite{RTBAS_2025,AEGIS_2026,jin2026understanding,lin2026mind}.
\ourapproach{} addresses cross-session skill composition by combining read-time provenance context restoration with pre-execution advisory review.
The system enforces context delivery and review while leaving the decision to revise, abandon, or confirm an operation to the task-execution agent, supporting safe artifact reuse.

\section{Conclusion}
\label{sec:conclusion}

In this paper, we study security risks in cross-session skill composition, showing how malicious influence from skill packages propagates through persistent task artifacts to affect decisions and operations in later sessions.
We characterize three attack patterns, Instruction Propagation, Data Boundary Violation, and Evidence Poisoning, and construct \bench{}, comprising 270 adversarial cases from 30 task scenarios across ten domains.
Across seven agent--model configurations, ASR ranges from 46.7\% to 77.4\%, demonstrating substantial risks in artifact handoffs within the evaluated workflows.
To support agents in safely reusing artifacts produced during skill execution, we propose \ourapproach{}, which combines provenance context restoration with pre-execution advisory review to provide evidence and revision advice while retaining the agent's final decision.
Across three agent configurations, it reduces overall ASR from 69.5\% to 10.9\% and ACSR from 54.8\% to 10.0\%, while increasing TSR from 79.6\% to 87.5\%.
These results show that helping agents identify and correct poisoned content can improve both security and legitimate task completion.
Safe cross-session artifact reuse requires distinguishing provenance, factual support, and user authorization, and reassessing artifact use before executing concrete operations.

% \section*{Data Availability}
% The experimental results are made publicly available in our artifact package~\cite{our_repo}.

% Bibliography (requires: references.bib in the same directory as main.tex)
\bibliographystyle{ACM-Reference-Format}
\bibliography{references}

% \appendix
% \input{Section/11-consequence-mapping}

\end{document}